# SANA -
# Security Analysis in Internet Traffic through Artificial Immune Systems

Michael Hilker[1] and Christoph Schommer[2]

[1] University of Luxembourg, Campus Kirchberg
1359, Luxembourg, 6, Rue Coudenhove-Kalergi, Luxembourg
`michael.hilker@uni.lu`
[2] University of Luxembourg, Campus Kirchberg
1359, Luxembourg, 6, Rue Coudenhove-Kalergi, Luxembourg
`christoph.schommer@uni.lu`

**Abstract.** The Attacks done by Viruses, Worms, Hackers, etc. are a
Network Security-Problem in many Organisations. Current Intrusion Detection Systems have significant Disadvantages, e.g. the need of plenty of
Computational Power or the Local Installation. Therefore, we introduce
a novel Framework for Network Security which is called SANA. SANA
contains an artificial Immune System with artificial Cells which perform
certain Tasks in order to to support existing systems to better secure
the Network against Intrusions. The Advantages of SANA are that it is
efficient, adaptive, autonomous, and massively-distributed. In this Article, we describe the Architecture of the artificial Immune System and the
Functionality of the Components. We explain briefly the Implementation
and discuss Results.

**Keywords.** Artificial Immune Systems, Network Security, Intrusion Detection, Artificial Cell Communication, Biological-Inspired Computing,
Complex Adaptive Systems

## 1 Introduction

Companies, Universities, and other Organisations use connected Computers,
Servers, etc. for Working, Storing of important Data, and Communication. These
Networks are an Aim for Attackers in order to breakdown the Network Service
or to gain internal and secret Information.

These Attacks are Intrusions which are e.g. Worms, Viruses, Hacker-Attacks.
Network Administrators try to secure the Network against these Intrusions using
Intrusion Detection Systems (IDS). The Network Intrusion Detection Systems
(NIDS) are a local System which is installed in one important Node and which
checks all Packets routed over this Node, e.g. SNORT [1] or [2,3,4,5,6]. Host-
based Intrusion Detection Systems (HIDS) are installed on each Node and check
each Packet which is routed over this Node [7,8,9]. Furthermore, there are approaches of distributed Intrusion Detection Systems (D-IDS) which install IDS
on all machines and connect these; one example is SNORTNET [10].

Unfortunately, these IDS have several Disadvantages as for example the plenty of Computational Power, the need of Administration during Execution, and local Installation. Additionally, the Intrusions are getting both more and more complex and intelligent, so that the IDS have lots of Problems to identify the Intrusions, e.g. Camouflage of Attacks. Thus, novel Approaches for Network Security are needed which should provide the following features:

– Distributed: all Nodes should be secured and there should not be any central Center
– Autonomous: the System and all Components should work autonomously; hereby, the number of false-positives should be low
– Adaptive: the System should have the ability to identify or react to modified or even novel Attacks
– Cooperative: The Computational Power should be shared over the whole Network

In SANA, we introduce an artificial Immune System which provides the features explained above. In the next Section, we discuss existing artificial Immune Systems for the Application of Network Security.

## 2   Current Situation

For the explanation of the different existing artificial Immune Systems for Network Security, we will introduce briefly the Paradigm of artificial Immune Systems [11]:

An artificial Immune System tries to simulate the human Immune System which secures the Human Body against Pathogens [12]. An artificial Immune System is a massively distributed System and Complex Adaptive System with lots of components. In the human Immune System, these Components are e.g. Cells, Lymph-Nodes, Bone Marrow. All of these Components work autonomously, efficiently and are highly specialised. These Components cooperate using the Cell Communication with e.g. Cytokines and Hormones. Additionally, there are lots of cellular and immunological Processes which mesh in the Protection of the Human Body. The artificial Immune Systems try to model these. Unfortunately, the human Immune System and the Modelling of it is so complex and partly not understood. Therefore, artificial Immune Systems can only model a part of the human Immune System.

There are several artificial Immune Systems for Network Security. We discuss some interesting Approaches of artificial Immune Systems for Network Security:

Spafford and Zamboni introduce in [13] a System for Intrusion Detection using autonomous Agents. These Agents cooperate with Transceivers and do not move through the Network. Hofmeyr and Forrester [14,15,16] introduce an artificial Immune System for Network Security (named ARTIS/LISYS). The AIS models the Lifecycle of T- and B-Cells with positive and negative Selection. The non-mobile Detectors check a Triple of Source-IP, Destination-IP and Destination-Port and evaluate if a Packet is malicious or not. Additionally, in this Broadcast-Network, all Detectors see all Packets and react to it. In [17] an artificial Immune System as a Multi-Agent System is introduced for Intrusion Detection. The system uses mobile Agents which cooperate with a centralised Database containing the Attack-Information.

In the next Section we introduce the Architecture of the artificial Immune System SANA. In contrast to the existing artificial Immune Systems, SANA uses autonomous, fully-mobile, and lightweighted artificial Cells; additionally, SANA does not have any centralised System. Furthermore, SANA is not a closed Framework; it is possible to use existing Network Security Approaches in SANA. Thereafter, we take a closer look on the different Components of the artificial Immune System.

## 3   SANA - Architecture

The artificial Immune System of SANA secures the whole Network against Intrusions and provides the Features explained above. In SANA, we simulate a packet-oriented Network using a Network Simulator (see Section 3.1). SANA is a collection of non-standard Approaches for Network Security and we test if they increase the Performance of existing Network Security Systems. An Adversarial injects Packets with and without Attacks in order to stress the Network and the artificial Immune System as well as to simulate Attacks (Section 3.2).

The artificial Immune System uses several Components for the Security of the Network. All of these Components work autonomously and there is no Center which is required by any Component. The main Components are artificial Cells, Packet-Filters, IDS, etc. Packet-Filters are a local System that check the Header of each Packet. IDS are local, non-mobile Systems which check Packets and observe the Network Traffic in order to secure the Node where the IDS is installed. Artificial Cells (Section 3.3) are autonomous, fully-mobile, and lightweighted Entities which flow through the Network and perform certain Tasks for Network Security, e.g. Packet-Checking, Identification, of Infected Nodes or Monitoring of the Network. Furthermore, artificial Cell Communication (Section 3.4) is used to initialise Cooperation and Collaboration between the artificial Cells and a Self-Management (Section 3.5) is utilised for a Regulation of the artificial Immune System. In the next Sections, we take a closer look on the different Components of SANA.

### 3.1   Network Simulator, Security Framework and Workflow

The Network Simulator simulates a Packet-Oriented Network and is based on the Adversarial Queueing Theory [18,19,20]. The Simulator uses a FIFO (First In First Out) approach for Queueing and for Routing the Shortest Path Routing with the Dijkstra-Algorithm. It has a Quality of Service (QoS) Management which prefers artificial Cells and other important Messages that are sent between certified Components of the AIS.

The Security Framework is the AIS which must be installed on each Node of the Network. Furthermore, this Framework guarantees e.g. the execution of the artificial Cells, the Presentation of Packets to all Security Components, the Sending of Messages. The Design of the Security Framework is focussed on Expandability in order to enhance it and to use existing Approaches in Network Security. One example of a Network Security Approach is Malfor [21], a system for Identification of the Processes which are involved in the Installation of an Intrusion.

The Workflow is that each Packet is checked in each Node by every Security Component - e.g. artificial Cells, Packet-Filters, and IDS - each Security Component can perform other Tasks - e.g. moving to other Nodes or sending Messages - and the Adversarial injects Packets into the Network.

### 3.2   Adversarial and Attacks

An Adversarial has the Function to Stress the Network and the AIS using Packets with and without Attacks; it has to keep in mind that the bandwidth of the connection is limited and that the queues have limited size. The Adversarial injects Packets without Attacks in order to simulate a real Network. The Packets with Attacks try to infect Nodes with Attacks; the infected Nodes then perform certain Tasks depending on the Attack, e.g. sending Packets with Attack to other Nodes. The Attack is an abstract Definition for all Intrusions in SANA. So, nearly all Intrusions can be modelled, e.g. Worms, Viruses, and Hacker-Attacks.

### 3.3   Artificial Cells

Artificial Cells are the main Component in the artificial Immune System of SANA. An artificial Cell is a highly specialised, autonomous and efficient Entity which flows through the Network and performs certain Tasks for Network Security. In the Cooperation and with the enormous Number of artificial Cells, the whole System adapts quickly to Attacks and even to modified and novel Attacks; the idea of Complex Adaptive Systems (CAS) or Massively-Distributed Systems.

Each artificial Cell has the Job to perform some certain Task:

- ANIMA for Intrusion Detection which is a type of artificial Cells for checking Packets whether they contain an Attack or not. Furthermore, it compresses the Information how to identify and how to proceed if an Attack is found in order to save Storage-Space and Computational Power. More Information about ANIMA-ID can be found in [22].
- AGNOSCO which is a type of artificial Cells for the Identification of Infected Nodes using artificial Ant Colonies. It is a distributed System which identifies the infected Nodes quickly and properly. More Information can be found in [23].
- Monitoring artificial Cell which flows through the Network and collects Information about the Status and send this back to some certain Component, e.g. the Administrator.

- Using the Expandability of SANA, it is easily possible to introduce novel artificial Cells. Thus, it is e.g. possible to introduce artificial Cells for Anomaly Detection or Checking of the Status of a Network Node.
- Additionally, it is possible to use existing Approaches for Network Security. With the Expandability of SANA, these Approaches can be used in an artificial Cell; examples are Systems for Intrusion- [22,24] or Anomaly-Detection Systems [25,26,27].

### 3.4    Artificial Cell Communication

The idea in Complex Adaptive System (CAS) is that the Components (here: artificial Cells) perform basic Tasks, are highly specialised and use basic Systems for Cooperation. Only by Cooperation and the high amount of these Components, the System is adaptive and reaches the goal (here: Network Security).

The whole Architecture in SANA is composed without any central System. Thus, the artificial Cell Communication cannot use a Central Management System like it is used in several Multi Agent Systems or Ad-Hoc Networks. We model partly the Cell Communication of the Human Body in order to build up Communication and, thereafter, Cooperation between artificial Cells.

We introduce the Term Receptor which is a Public-Key-Pair. Each Component has Receptors and each Message is packed into a Substance which is an encrypted Message with Receptors. Only if a Receiver has the right Set of Receptors, it will receive the Message - the Idea of a Public-Key Infrastructure and widely used in Multi Agent System for the Disarming of Bad-Agents/-artificial Cells; however, in our Implementation, there is not any centralised Key-Server.

Additionally, we introduce artificial Lymph Nodes and Central Nativity and Training Stations (CNTS). Artificial Lymph Nodes supply the artificial Cells with e.g. Knowledge, initiate other artificial Cells if an event occurs and artificial Lymph Nodes care about the Routing of Substances. CNTS train and release new artificial Cell in order to have an evolutionary Set of artificial Cells which are up-to-date. Both, artificial Lymph Nodes and Central Nativity and Training Stations, are redundant installed in the System.

### 3.5    Self-Management of the artificial Immune System

The Self-Management of the System is currently only rudimentary. The artificial Cells are autonomous and thus they flow through the Network and perform certain Tasks. However, one Problem of Massively-Distributed Systems or Complex Adaptive Systems is that they just do their Tasks but there is not any guarantee that the Systems will do the Tasks successfully. On the basis of the artificial Cell Communication and novel Structures, we want to introduce a distributed Self-Management of the artificial Immune System in order to give a certain amount of Guarantee. However, this is one of the Next Steps explained in the Section 6.

## 4  SANA - Implementation

The Project SANA is implemented in Java. The Network Simulator, Adversarial, and the artificial Immune System are implemented and running. Different Types of artificial Cells are implemented. The Performance of these artificial Cells is tested and they perform the Tasks properly. Attack-Scenarios are additionally implemented for Testing Purposes and one example is a realistic Worm-Attack which will be discussed in the Section 5.1.

The whole Implementation has the aim to give a Prototype for Testing and Evaluation of the Approaches. Furthermore, the Implementation focuses more on Expandability than on Performance; it is also possible to model nearly all Intrusions and nearly all immunological Processes. It is also possible to add common used Network Security Solutions like SNORT [1] or Malfor [21]. With this, we can compare the Performance of SANA with common used IDS and we can model cooperation between SANA and IDS.

## 5  SANA - Results

The Results we gained are promising. SANA identifies most Attacks - about 60%-85% - depending on the Attack-Behaviour, the Network Topology and the Behaviour of the artificial Immune System with the artificial Cells. The infected Nodes are identified quickly by AGNOSCO and the System adapts to Attacks using local Immunization.

If there are IDS or especially NIDS in the Network which protect important Nodes like the Internet Gateway or the E-Mail-Server, there is cooperation between SANA and the IDS with a good performance - about 80%-95% of the Attack are prevented. Thus, SANA does not replace existing IDS, it enhances them.

In the next Section, we discuss the Results of a Simulation of a realistic Worm-Attack.

### 5.1  Simulation of a Worm-Attack

In this Section, we discuss a Modelling of a realistic Worm-Attack onto the Network. The Worm enters a Network and uses a Security-Hole in a Node in order to install itself. After this, the Worm tries to propagate it to other Nodes; therefore, it sends lots of Packets containing a copy of it to other Nodes. SANA tries to identify and remove these Packets, identifies the infected Nodes and disinfects the identified infected Nodes. Therefore, SANA uses the different types of artificial Cells explained in the Section 3.3 and the artificial Cell Communication explained in the Section 3.4.

The Performance of SANA in this Simulation is promising. It secures other Nodes from being infected by this Worm using ANIMA for Intrusion Detection [22]; only some Neighbour-Nodes are infected (about 2-5 Nodes for each Infection). It also identifies the infected Nodes using AGNOSCO [23] quickly (about 50-150 Time-Steps for each infected Node) and using the artificial Cell Communication (Section 3.4), AGNOSCO informs the artificial Lymph-Nodes (Section 3.4) which start an artificial Cell for Disinfection which disinfect the Node fast. To sum up, SANA protects the Network against a Worm-Attack properly.

## 5.2   Theoretical Analysis of distributed IDS

In the theoretical Part of the SANA-Project, we compare the Performance and the Need of Resource of distributed and centralised Network Security Systems. Examples for centralised are e.g. IDS and for distributed AIS. However, the Analysis shows quickly that the Performance of the both Approaches is highly dependent on the Network Topology and the Behaviour of the Intrusions. The Analysis fortunately shows that the Performance of IDS is increased if AIS are added and the additionally needed Resources are limited.

## 6   SANA - Next Steps

Next Steps in the SANA-Project are to simulate realistic Attacks on Networks, e.g. different Worm, Virus and Malwar-Attacks; also Attacks which consists of several different Attacks. Additionally, another part is to increase the Performance of the artificial Cell Communication (Section 3.4) and analyse the Performance of it theoretically. Furthmore, we will introduce a Self-Management (Section 3.5) which guarantees a certain amount of Security and we will perform further theoretical Comparison (Section 5.2) between distributed and centralised Network Security Systems.

## 7   Conclusion

Network Security is still a challenging field. Unfortunately, the Attacks are getting both more complex and intelligent. Therefore, existing Network Security Systems have problems to cope with these Problems. We introduce with SANA an artificial Immune System with several non-standard Approaches for Network Security. With the gained Results, we are sure that SANA will enhance current Network Security Systems.

One last word about SANA: SANA is Latin and stands for healthy. Furthermore, the Work is done interdisciplinary in cooperation between Researchers from Biology and Computer Science.

## Acknowledgments

The PhD-Project SANA is part of the project INTRA (= INternet TRAffic management and analysis) that are financially supported by the University of Luxembourg. We would like to thank the Ministre Luxembourgeois de l'education et de la recherche for additional financial support.

# References

1. Roesch, M.: Snort - lightweight intrusion detection for networks. LISA **13** (1999) 229–238
2. Debar, H., Dacier, M., Wespi, A.: Towards a taxonomy of intrusion-detection systems. Computer Networks **31** (1998) 805–822
3. Snapp, S.R., Brentano, J., Dias, G.V., Goan, T.L., Heberlein, L.T., lin Ho, C., Levitt, K.N., Mukherjee, B., Smaha, S.E., Grance, T., Teal, D.M., Mansur, D.: DIDS (distributed intrusion detection system) - motivation, architecture, and an early prototype. National Computer Security Conference **14** (1991) 167–176
4. Staniford-Chen, S., Cheung, S., Crawford, R., Dilger, M., Frank, J., Hoagland, J., Levitt, K., Wee, C., Yip, R., Zerkle, D.: Grids - a graph based intrusion detection system for large networks. National Information Systems Security Conference **19** (1996)
5. Janakiraman, R., Waldvogel, M., Zhang, Q.: Indra: A peer-to-peer approach to network intrusion detection and prevention. Proceedings of IEEE WETICE 2003 (2003)
6. Antonatos, S., Anagnostakis, K., Polychronakis, M., Markatos, E.: Performance analysis of content matching intrusion detection systems. SAINT **4** (2004)
7. Wagner, D., Dean, D.: Intrusion detection via static analysis. In IEEE Symposium on Security and Privacy (2001)
8. Lindqvist, U., Porras, P.A.: expert-bsm: A host-based intrusion detection solution for sun solaris. In Proceedings of the 17th Annual Computer Security Applications Conference (2001) 240–251
9. Chari, S.N., Cheng, P.C.: Bluebox: A policy-driven, host-based intrusion detection system. ACM Transactions on Information and System Security **6** (2003) 173–200
10. Fyodor, Y.: Snortnet' - a distributed intrusion detection system. [Online]. Available: http://snortnet.scorpions.net/snortnet.pdf (2000)
11. DeCastro, L.N.: Artificial Immune Systems: A New Computational Intelligence Approach. First edn. Springer (2002)
12. Janeway, C.A., Travers, P., Walport, M., Shlomchik, M.: Immunobiology: the Immune System in Health and Disease. Sixth edn. Garland Publishing (2004)
13. Spafford, E.H., Zamboni, D.: Intrusion detection using autonomous agents. Computer Networks **34** (2000) 547–570
14. Hofmeyr, S.A., Forrest, S.: Immunity by design: An artificial immune system. Proceedings of the Genetic and Evolutionary Computation Conference **2** (1999) 1289–1296
15. Hofmeyr, S.A., Forrest, S.: Architecture for an artificial immune system. Evolutionary Computation **8** (2000) 443–473
16. Hofmeyr, S.A., Forrest, S.: Immunology as information processing. (2000)
17. Machado, R.B., Boukerche, A., Sobral, J.B.M., Juca, K.R.L., Notare, M.S.M.A.: A hybrid artificial immune and mobile agent intrusion detection based model for computer network operations. IPDPS '05: Proceedings of the 19th IEEE International Parallel and Distributed Processing Symposium (IPDPS'05) - Workshop 6 **19** (2005)
18. Andrews, Baruch Awerbuch, Antonio Fernndez, Tom Leighton, Zhiyoung Liu and Jon Kleinberg, M.: Universal-Stability Results and Performance Bounds for Greedy Contention-Resolution Protocols. Journal of the ACM **48** (2000) 39–69
19. Hilker, M.: Queueing Strategies in Internet Routing. Diploma Thesis at the Johann Wolfgang Goethe-University Frankfurt/M., Germany (2005)

20. Hilker, M., Schommer, C.: A new queueing strategy for the adversarial queueing theory. IPSI-2005 Slovenia (2005)
21. Neuhaus, S., Zeller, A.: Isolating intrusions by automatic experiments. 13th Annual Network and Distributed System Security Symposium (2006)
22. Hilker, M., Schommer, C.: Description of bad-signatures for network intrusion detection. AISW-NetSec 2006 during ACSW 2006, CRPIT **54** (2006)
23. Hilker, M., Schommer, C.: Agnosco - identification of infected nodes with artificial ant colonies. RASC 2006 (2006)
24. Finizio, I., Mazzariello, C., Sansone, C.: A temporal-behavior knowledge space for detecting intrusions in computer networks. RASC 2006 (2006)
25. Sekar, R., Gupta, A., Frullo, J., Shanbhag, T., Tiwari, A., Yang, H., Zhou, S.: Specification-based anomaly detection: a new approach for detecting network intrusions. Volume 9. (2002) 265–274
26. Lazarevic, A., Ertoz, L., Ozgur, A., Srivastava, J., Kumar, V.: A comparative study of anomaly detection schemes in network intrusion detection. Proceedings of Third SIAM Conference on Data Mining **3** (2003)
27. Leung, K., Leckie, C.: Unsupervised anomaly detection in network intrusion detection using clusters. Australasian Computer Science Conference **28** (2005)